\documentclass[conference,compsoc]{IEEEtran}
\usepackage{graphicx}
\usepackage{hyperref}
\usepackage{cite}
\usepackage{url}
\usepackage[utf8]{inputenc}
\usepackage{multirow}
\usepackage{filecontents}
\usepackage{amsmath,amssymb,amsfonts}
\usepackage{xcolor}
\usepackage{tcolorbox}
\newcommand{\fixme}[1]{{\color{black} #1}}

\begin{document}
%

\title{Exploring Parent-Child Perceptions on Safety in Generative AI: Concerns, Mitigation Strategies, and Design Implications }



%
\author{\IEEEauthorblockN{Yaman Yu\IEEEauthorrefmark{1},
Tanusree Sharma\IEEEauthorrefmark{2},
Melinda Hu\IEEEauthorrefmark{3}, 
Justin Wang\IEEEauthorrefmark{4},
Yang Wang\IEEEauthorrefmark{1}}

\IEEEauthorblockA{\IEEEauthorrefmark{1}
University of Illinois at Urbana-Champaign,\\ 
{yamanyu2, yvw}@illinois.edu}

\IEEEauthorblockA{\IEEEauthorrefmark{2} Pennsylvania State University\\tfs5747@psu.edu}
\IEEEauthorblockA{\IEEEauthorrefmark{3}The Hockaday School\\melindamhu@gmail.com}

\IEEEauthorblockA{\IEEEauthorrefmark{4}Dougherty Valley High School\\ justin.wangjw1@gmail.com}}



\maketitle

\begin{abstract}
The widespread use of Generative Artificial Intelligence (GAI) among teenagers has led to significant misuse and safety concerns. To identify risks and understand parental controls challenges, we conducted a content analysis on Reddit and interviewed 20 participants (seven teenagers and 13 parents). Our study reveals a significant gap in parental awareness of the extensive ways children use GAI, such as interacting with character-based chatbots for emotional support or engaging in virtual relationships. Parents and children report differing perceptions of risks associated with GAI. Parents primarily express concerns about data collection, misinformation, and exposure to inappropriate content. In contrast, teenagers are more concerned about becoming addicted to virtual relationships with GAI, the potential misuse of GAI to spread harmful content in social groups, and the invasion of privacy due to unauthorized use of their personal data in GAI applications. The absence of parental control features on GAI platforms forces parents to rely on system-built controls, manually check histories, share accounts, and engage in active mediation. Despite these efforts, parents struggle to grasp the full spectrum of GAI-related risks and to perform effective real-time monitoring, mediation, and education. We provide design recommendations to improve parent-child communication and enhance the safety of GAI use.

\end{abstract}


%

\vspace{-2mm}
\section{Introduction}
\vspace{-2mm}
\label{intro}

Generative artificial intelligence (GAI) has swiftly become a vital component of the digital ecosystem, significantly transforming user interactions with technology and opening up new avenues for creativity and innovation. Examples like ChatGPT, which enhances content creation with advanced text generation, and DALL-E, which creates realistic images from text descriptions, illustrate GAI's impact on digital media and design. \fixme{In contrast, Character.ai offers a unique experience by enabling users to engage with character-based AI chatbots, allowing for more personalized and interactive conversations, often mimicking human-like interactions.}
This widespread adoption is particularly evident among younger age groups. A study by the UK’s communications regulator Ofcom finds that 79\% of online 13-17 year-olds and 40\% of 7-12 year-olds in the UK are using generative AI tools and services~\cite{ofcom2023}. As young generations rapidly adopt emerging technologies, it is crucial to consider the potential risks and harms these technologies may bring, including concerns related to privacy, and safety. Particularly, the usage of GAI by teens and young adults is closely linked to their mental health~\cite{usnews2024}. Therefore, it is imperative to thoroughly understand their usage patterns and evaluate the current protective measures. By doing so, we can develop and implement more effective precautions to safeguard their well-being as they engage with these technologies.

Previous studies have examined the risk perceptions and parental controls across various technologies and applications, including Virtual Reality~\cite{cao2024understanding}, IoT devices~\cite{sun2021child}, social media platforms~\cite{alsoubai2024profiling}, and gaming~\cite{faraz2022child}. These studies identified both common and unique risks in various contexts, including explicit content, addiction, cyberbullying, and harassment~\cite{jones2012trends, mesch2009parental, livingstone2011risks, livingstone2008parental}. Subsequent research has also highlighted parents' mediation strategies, such as monitoring, restriction, and active mediation~\cite{wisniewski2017parental}. Additionally, studies have pointed out the trade-offs that exist between parental control and children's autonomy. Since children can benefit from low-risk experiences to develop their ability to identify and manage risks independently in the future, defining the boundaries of parental control and ensuring the right balance remain challenging in children's risk management~\cite{wisniewski2015resilience, wang2022informing}. The motivation for our research stems from the need to better understand both parents' and children's safety perceptions and coping strategies within the context of Generative AI (GAI), an emerging technology with unique risks and opportunities. We focus on three research questions:

\begin{itemize}
    \item RQ1: How did teenagers use GAI?
    \item RQ2: What are the perceptions of parents and children regarding safety with GAI?
    \item RQ3: What are the mediation strategies employed by parents to ensure children's safety with GAI, and what challenges do they face?
\end{itemize}

\fixme{To address these questions, we utilized a mixed-methods approach that combined Reddit content analysis, followed by semi-structured interviews involving teenagers and parents. Initially, we collected 712 posts and 8,533 comments from relevant subreddits, providing naturalistic insights into how teenagers engage with GAI. This was followed by semi-structured interviews with seven teenagers and 13 parents, designed to explore their in-depth perspectives on GAI use, safety concerns, privacy issues, and the strategies parents employ to mediate their teenagers’ interactions with these technologies. }

\textbf{Findings.} For RQ1, our findings reveal that child participants use Generative AI (GAI) in diverse and unexpected ways, far beyond the educational or experimental purposes that parent participants assume. Interviews with child participants and analysis of data from Reddit discussions indicate that teenagers often integrate GAI into their daily routines. They use GAI as an friend or confidant, providing emotional support and companionship. Additionally, teenagers reported using GAI to enhance their social interactions, such as incorporating bots into group chats to facilitate conversations and simulate social dynamics. Our research identified a primary reason for this gap between parents and children: the diverse sources from which children are informed and introduced to AI. Besides their parents, children learn about GAI through influencer videos, targeted advertisements on social media, and peer recommendations. These sources often introduce them to new and experimental GAI platforms, which typically carry more risks compared to more established ones. Furthermore, both parents and children hold significant misconceptions about GAI's data collection, sharing, and processing practices. \fixme{Teenagers often view GAI as either functioning like a search engine or a vast database, believing it directly pulls responses from the internet or provides answers from a stored set of responses. Similarly, many parents shared this search engine misconception, but some parents also incorrectly assumed that GAI platforms perform fact-checking and verification before generating responses.}\\
For RQ2, we uncovered significant security and privacy concerns associated with children's use of generative artificial intelligence (GAI), which parents frequently overlook due to unawareness of the varied interactions children have with these technologies. Child participants expressed specific worries about GAI in social contexts. They were concerned about becoming overly dependent on virtual relationships with chatbots, which could exacerbate their challenges in forming real-life social connections. Additionally, they noted the potential misuse of GAI in group chats to create distressing or harassing content. Moreover, child participants shared concerns regarding the unauthorized use of their personal data to create GAI chatbots or memes, and the risk of strangers misusing their online images to glean personal information. They also encountered not safe for work (NSFW) chatbots promoted by peers and social media, and some even created chatbots to spread harmful content, such as racist remarks. Parents, however, often underestimated these risks. While they were aware of GAI collecting basic information like names and locations, they did not fully appreciate the extent of sensitive data their children might share with GAI perceived as friends or confidants. This data could include details of personal traumas, medical records, and private aspects of their social and sexual lives.\\
For RQ3, leading GAI platforms, such as ChatGPT and Meta AI, offer limited protections, primarily focusing on banning explicit content and requiring users to be 13 or older. The absence of comprehensive parental control features on these platforms compels parents to rely on often ineffective tools for real-time mediation. These include using general parental control tools like iOS Family Control and Google Family Link, engaging in active mediation through open communication and setting consensual rules, and monitoring their children's GAI usage by sharing accounts or checking interaction histories. However, these strategies face significant challenges, such as the difficulty in finding clear information on GAI capabilities and restrictions, particularly concerning children's protection. Parents also find it challenging to monitor, adjust, and educate their children on GAI usage in real-time to prevent low-risk behaviors from escalating.\\

\textbf{Contributions.} This paper offers three main contributions to the field. First, we identified unique usage patterns and associated risks of GAI among teenagers. Second, we investigated the existing GAI mediation strategies employed by parents and identified challenges they face. Third, we evaluated both children's and parents' perspectives on desired GAI mediation and derived constructive design implications for future GAI systems. \fixme{Specifically, we recommend that GAI platforms incorporate age-appropriate content moderation, customizable parental control tools, and transparent risk disclosures to better address the safety needs of teenage users.}

\section{Related Work}

\vspace{-2mm}
\subsection{Children digital safety concerns}
\vspace{-2mm}
Security and privacy are crucial aspects of users' interaction with technology. Threat models and taxonomies provide a structured way to identify and categorize potential security threats and vulnerabilities, helping to understand and mitigate risks systematically across different technologies and scenarios~\cite{solove2005taxonomy, shostack2014threat, myagmar2005threat}. While comprehensive security and privacy threat models and taxonomies have been explored by researchers, user-perceived safety risks and their precautions vary based on their own background and experiences~\cite{anell2020end}. Thus, other studies have also explored usable security and privacy among diverse populations~\cite{kaushik2023guardlens, yu2023design, kaushik2024cross, sharma2023user, yu2024don, frik2019privacy}, such as people with disabilities, old adults, those from low socioeconomic backgrounds and children. Children specifically face unique challenges and risks in the digital realm because they are still developing their critical thinking and decision-making skills, making them more vulnerable to manipulation and exploitation. They may be more susceptible to cyberbullying, online harassment, data breaches, and exposure to inappropriate content~\cite{jones2012trends, mesch2009parental, livingstone2011risks, livingstone2008parental} . Additionally, they are often less aware of privacy issues and the long-term consequences of sharing personal information online~\cite{kumar2023understanding}.

To address these issues, researchers need to answer fundamental questions such as: What risks are children facing? What factors influence their behavior in relation to these risks and potential harms? What measures can be implemented to prevent these harms? Many researchers have summarized adolescent online risk exposure and their coping strategies by collecting self-reported data points~\cite{cranor2014parents, livingstone2006drawing, livingstone2008taking, wisniewski2014adolescent, cao2024understanding}. These research highlighted different categories of risks that teenagers encounter in various contexts. Some risks, such as exposure to explicit content, are common across all areas~\cite{livingstone2008taking, cao2024understanding}. However, certain safety concerns are unique to specific technologies or scenarios. For example, Virtual Reality introduces more vivid and unique ways of socializing with strangers, leading to specific security and privacy concerns in social interactions for children~\cite{cao2024understanding}. 

Generative AI is prevalently used in daily life and is widely accepted as a potential revolution for the future. Researchers have identified several security and privacy concerns for general users of Generative AI, such as disclosing sensitive information~\cite{zhang2024s} and the new privacy risks created by AI~\cite{lee2024deepfakes}. However, the specific risks that children experience with Generative AI tools or AI-powered platforms and how they cope with these risks remain largely unexplored in scholarly work. Understanding the unique challenges faced by children in this context is crucial for developing effective strategies to protect them. To address this gap, we are combining naturalistic data from Reddit posts and self-reported data from user interviews to comprehensively investigate the risks children face with Generative AI.

\vspace{-2mm}
\subsection{Mitigation strategies for children digital safety}
\vspace{-2mm}
Ensuring children's digital safety is a multifaceted challenge, rooted in their unique developmental stages and the complex social dynamics influencing security and privacy decisions~\cite{cranor2014parents, livingstone2008taking, wisniewski2014adolescent}. Parents are often seen as the primary gatekeepers responsible for safeguarding their children in the digital realm~\cite{sun2021they, shmueli2010privacy}. Wisniewski et al. outlined a framework for parental strategies to enhance teen online safety, known as the Teen Online Safety Strategies (TOSS)~\cite{wisniewski2017parental}. This framework includes three primary approaches: (1) observing children's online behavior without direct intervention (monitoring), (2) setting rules and limitations on internet use to regulate their activities (restriction), and (3) engaging in proactive discussions with children to educate them about safe online practices and encourage open communication (active mediation).

Many researchers have worked on monitoring and restriction to build up effective tools in various scenarios~\cite{iftikhar2021designing}. Most tools designed to enhance children's online safety primarily focus on monitoring or automatically controlling content, often neglecting children's agency and their perspectives on risks and coping strategies~\cite{al2019dnsbl, chiu2019defense, kumar2021parental}. Few studies explored the family risk management solutions, including open communication, self-regulation, and granular controls~\cite{dumaru2024s, akter2022parental}. However, researchers have highlighted the tension between children's autonomy and parental control in managing online risks. Children and parents hold different perceptions on security and privacy~\cite{cranor2014parents}. 

Additionally, children desire privacy while still receiving protection from their parents~\cite{stewart2022parental}. Overrestrictive and invasive parental control methods are perceived extremely negatively by children and can negatively impact family relationships~\cite{ghosh2018safety, wisniewski2017parental}. These methods can lead to a lack of trust and communication between parents and children, undermining the intended protective benefits~\cite{wang2021protection}. Wisniewski et al. conducted a longitudinal diary study to gain deep and contextualized knowledge of teens' risk experiences and coping strategies within various risk categories~\cite{wisniewski2016dear}. They found that most of the risks teenagers reported were lower-risk online situations. Exposure to these lower-risk scenarios can help teens build resilience and develop interpersonal skills, which can mitigate future harm. Resilience has been identified as a key factor in protecting teens from experiencing significant online risks~\cite{wisniewski2015resilience}. Therefore, finding the right balance in parental control and children's self-protection is critical to ensuring adolescent safety as well as their moral, social, and emotional development~\cite{wisniewski2015preventative}. However, parents' and children's perceptions of Generative AI risks and their security and privacy practices remain unexplored in current research. To address this gap, we aim to conduct in-depth user interviews to identify the pros and cons of their current practices and stimulate discussion on appropriate mediation strategies for children's use of Generative AI.
\vspace{-2mm}

\section{Method}
\vspace{-2mm}
\subsection{Reddit Study}
\vspace{-2mm}
We first collected 712 posts and 8,533 comments on April 9, 2024, using the Python Reddit API Wrapper (PRAW)~\footnote{https://praw.readthedocs.io/en/stable/}. We gathered the data from various relevant subreddits, ensuring a broad and comprehensive understanding of how children interact with Generative AI platforms. Through qualitative analysis of this Reddit data, we were able to uncover detailed insights into the rich and prevalent usage of Generative AI platforms by children. This analysis highlighted the diverse ways in which children engage with these technologies, as well as the potential challenges and risks they encounter. These findings 
provide a solid foundation to explore these themes further in subsequent in-depth interviews.

\subsubsection{Data Collection}
 To comprehensively cover content related to our research questions on teenagers and Generative AI, we first created a list of search keywords by identifying close terminologies related to \textit{``teenager''} (general keywords) and \textit{``Generative AI''} (technology-focused keywords). \fixme{We utilized a combination of general and technology-focused keywords in our search. We employed general terms such as teen, teenager, adolescent, and high school student. These keywords were designed to capture posts authored by or discussing teenagers. For the technology focus, we used terms such as Generative AI, AI chatbot, artificial intelligence, AI, ChatGPT, DALL-E, and Midjourney. These keywords targeted discussions specifically about the use of popular AI tools and platforms. We conducted open searches combining these keywords across Reddit to gather data from various subreddits.} Other than open searches, we also applied specific criteria to select subreddits, ensuring comprehensive coverage of relevant discussions: these subreddits should focus either on the teenager community or Generative AI technology. We chose subreddits with the most active users online during our browsing sessions. The full list of subreddits and search keywords used is detailed in Table~\ref{tab:subreddit}.

We first performed open searches combining these keywords across the entire Reddit platform. Then, we searched for all terms related to \textit{``teenager''} within Generative AI subreddits (r/midjourney, r/ChatGPT, r/OpenAI), and all terms related to \textit{``Generative AI''} within teenager subreddits (r/teenagers, r/AskTeenGirls, r/askteenboys, r/BisexualTeens). Finally, we removed any duplicates from the search results. The number of posts collected from each subreddit is detailed in Table~\ref{tab:subreddit}.
 
\subsubsection{Analysis}
Three researchers reviewed each post to filter those posted by teenagers and related to GAI topics. Two criteria were used to verify if a post was authored by a child: (1) the age disclosed near the username of each post on the Reddit platform in certain subreddits, and (2) language indicative of a teenager, such as mentions of being in high school, stating their age (e.g., \textit{``I am 13.''}), or referencing parental guidance (e.g., \textit{``my parents told me....''}). The research team categorized related posts or comments into five overarching high-level themes: \textit{“Teenager GAI Usage Patterns”}, \textit{“Teenager GAI Attitudes”}, \textit{“Teenager Understanding of GAI”}, \textit{“Teenager GAI Benefits”}, and \textit{“Teenager GAI Concerns”}. Within these categories, 53 level-2 themes were identified, such as \textit{“Using GAI chatbot as confidant”} and \textit{“GAI replacing human labor”}. During the analysis process, researchers regularly convene to discuss discrepancies and emerging themes in the codebook, aiming to reach a consensus. These categories allowed us to investigate RQ1 and partially address RQ2. We further reported the findings on children's GAI usage and their concerns in Section~\ref{Reddit Analysis Results}.


\begin{table*}[]
\centering
\renewcommand{\arraystretch}{1.2}
\begin{tabular}{clccc}
\hline
\multicolumn{2}{c}{\textbf{Subreddit}} &
  \textbf{\# of Posts Pulled} &
  \textbf{\# of Comments Pulled} &
  \textbf{\# of Related Posts/Comments} \\ \hline
\multicolumn{1}{l|}{} &
  \begin{tabular}[c]{@{}l@{}}Open search\\ all subreddits\end{tabular} &
  159 &
  2034 &
  32 \\ \hline
\multicolumn{1}{c|}{\multirow{4}{*}{teenager}} & r/teenagers   & 243 & 4294 & 90 \\ \cline{2-5} 
\multicolumn{1}{c|}{}                          & r/BisexualTeens     & 45  & 203  & 24 \\ \cline{2-5} 
\multicolumn{1}{c|}{}                          & r/askteenboys     & 10  & 18   & 7  \\ \cline{2-5} 
\multicolumn{1}{c|}{}                          & r/AskTeenGirls    & 7   & 15   & 11 \\ \hline
\multicolumn{1}{c|}{\multirow{3}{*}{GAI}}      & r/ChatGPT    & 158 & 1783 & 10 \\ \cline{2-5} 
\multicolumn{1}{c|}{}                          & r/OpenAI     & 9   & 34   & 7  \\ \cline{2-5} 
\multicolumn{1}{c|}{}                          & r/midjourney & 81  & 152  & 0  \\ \hline
\end{tabular}
\vspace{0.2cm}
\caption{Overview of statistics in the Reddit dataset.}
\label{tab:subreddit}
\end{table*}

\vspace{-2mm}
\subsection{Interview Study}
\vspace{-2mm}
In addition to our Reddit study, we conducted a semi-structured interview study between January and May 2024. We interviewed 7 children and 13 parents to gain a deeper understanding of their experiences and risk perceptions related to Generalized Artificial Intelligence (GAI). We also aimed to identify unmet needs for parental control features in GAI tools. We developed two tailored versions of a user interview protocol, one for parents and one for children, aligned with our research questions. The protocol was structured according to the following topical sections: (1) current practice-related parental mediation strategies to ensure privacy, security, and safety for teenagers; (2) current use of GAI tools by both parents and children and their perceptions towards the risk of GAI tools (3) existing parental control framework or design fits the GAI tool scenario in protecting children's security, privacy and safety. The questions in the children's interview protocol were tailored to accommodate different age groups, ensuring developmental appropriateness and clarity, and thereby enabling us to gather rich and meaningful insights from both children and parents. These interviews were typically around one hour in length.

\subsubsection{Participant Recruitment}


\fixme{We recruited parent participants through Prolific based on the following inclusion criteria: (1) prior experience with Generative AI, including the duration of use and the specific platforms they have engaged with
, (2) residing in the U.S. and fluent in English, and (3) having at least one child aged 13-17. Teenager participants were recruited from local high schools and through Prolific, with some parents introducing their own children as well. This approach ensured diversity in our participant pool, with variations in both the platforms used and the duration of their Generative AI experience.}


(i) To recruit teenagers, we sent out a student survey to a local public high school in the US. 
If students expressed interest in participating, we obtained written informed consent from parents via email, following our institution's approved protocols and procedures for securing consent; (ii) We also recruited teenagers through Prolific by distributing a parent survey. Parents who expressed interest in participating will be interviewed first; then, we will provide the opportunity to invite their children to participate in our study as wells. If parents expressed interest in having their children participate in the study and provided informed consent, we shared the student survey with them via email. All participants who completed the interviews were compensated with a \$25 Amazon gift card. The demographics of our participants are summarized in Table~\ref{tab:demo}.

\begin{table*}[]
\centering
\renewcommand{\arraystretch}{1.5}
\setlength{\tabcolsep}{4mm}{
\begin{tabular}{cccccl}
\hline
\textbf{ID} & \textbf{Age} & \textbf{Gender} & \textbf{Used GAI} & \textcolor{black}{\textbf{Months of Use}} & \textcolor{black}{\textbf{State}} \\ \hline       
P1  & 17 (teen)  & Male   & ChatGPT                                  & \textcolor{black}{1 to 3 months} & \textcolor{black}{IL} \\
P2  & 14 (teen)  & Male   & ChatGPT, Bard, Character.ai              & \textcolor{black}{1 to 3 months} & \textcolor{black}{IL} \\
P3  & 16 (teen)  & Male   & ChatGPT, CoPilot, Logo ai generator      & \textcolor{black}{6 months to 1 year} & \textcolor{black}{IL} \\
P4  & 13 (teen)  & Female & ChatGPT                                  & \textcolor{black}{1 to 3 months} & \textcolor{black}{NY} \\
P5  & 15 (teen)  & Male   & ChatGPT, Midjourney, Deepfake Voice      & \textcolor{black}{6 months to 1 year} & \textcolor{black}{IL} \\
P6  & 13 (teen)  & Male   & ChatGPT, Character.ai                    & \textcolor{black}{Less than 1 month} & \textcolor{black}{OH} \\
P7  & 15 (teen)  & Male   & DALL-E, Character.ai, Chai AI            & \textcolor{black}{6 months to 1 year} & \textcolor{black}{VA} \\
P8  & 54  & Male   & ChatGPT                                  & \textcolor{black}{3 to 6 months} & \textcolor{black}{DC} \\
P9  & 37  & Male   & ChatGPT                                  & \textcolor{black}{3 to 6 months} & \textcolor{black}{CO} \\
P10 & 34  & Female & ChatGPT                                  & \textcolor{black}{6 months to 1 year} & \textcolor{black}{TX} \\
P11 & 52  & Female & ChatGPT                                  & \textcolor{black}{1 to 3 months} & \textcolor{black}{NY} \\
P12 & 44  & Male   & ChatGPT                                  & \textcolor{black}{3 to 6 months} & \textcolor{black}{OH} \\
P13 & 35  & Female & ChatGPT                                  & \textcolor{black}{3 to 6 months} & \textcolor{black}{MI} \\
P14 & 49  & Female & ChatGPT, DALL-E, Midjourney              & \textcolor{black}{6 months to 1 year} & \textcolor{black}{VA} \\
P15 & 35  & Female & ChatGPT                                  & \textcolor{black}{1 to 3 months} & \textcolor{black}{VA} \\
P16 & 37  & Female & ChatGPT                                  & \textcolor{black}{3 to 6 months} & \textcolor{black}{SC} \\
P17 & 36  & Male   & ChatGPT, DALL-E, Midjourney              & \textcolor{black}{More than 1 year} & \textcolor{black}{IN} \\
P18 & 42  & Male   & ChatGPT, DALL-E, Midjourney              & \textcolor{black}{6 months to 1 year} & \textcolor{black}{NV} \\
P19 & 43  & Female & ChatGPT, DALL-E, Diffit, Magic School AI & \textcolor{black}{3 to 6 months} & \textcolor{black}{GA} \\
P20 & 35  & Female & ChatGPT, DALL-E                          & \textcolor{black}{3 to 6 months} & \textcolor{black}{FL} \\ \hline
\end{tabular}
}
\vspace{0.2cm}
\caption{Participants demographics with usage patterns and locations highlighted.}
\label{tab:demo}
\end{table*}
\subsubsection{Interview Procedure}
\vspace{-2mm}
\hfill\\
We emailed parent participants one day before their scheduled interview to obtain their written consent. Similarly, for teenagers, we sent emails to both the teenagers and their parents to secure consent from both parties. We conducted semi-structured interviews with each participant via Zoom, where we obtained verbal consent from all participants before recording the sessions. Each interview was audio recorded and transcribed verbatim. To address our research questions, each interview was divided into three main sections: initially, we discussed family background and the parental controls in place; secondly, we explored participants' experiences with GAI and their perceptions of associated risks; finally, we delved into their strategies and challenges in mediating GAI usage.

\subsubsection{Interview protocol}
\label{sec:interview}
\hfill\\
\textbf{Interview with Teenagers.} For the interviews with teenagers, the first section focused on gathering background information about their living environment, including the number of people in their household, device access, and the type of devices they use. This helped us understand if they share devices and with whom. We asked participants, \textit{``Do you personally own these devices or do you share them with other family members? If you share, can you specify with whom?''}
We then inquired about any parental controls implemented on their devices or technologies, asking, \textit{``Have they set up any rules or restrictions for your device/technology usage, such as for gaming, social media, or any apps/websites you use?''} We also asked about how often their parents monitor the devices and how well they understand these functionalities. Finally, we asked about their feelings regarding the current parental controls.

In the next section, we focused on their current usage and understanding of generative AI (GAI) tools. We started by discussing their specific use of GAI and what they like about these applications. We encouraged them to imagine and describe how GAI works, for example, by asking, \textit{``Can you create a drawing or describe how you imagine tools like ChatGPT work?''} This helped us gauge their level of understanding. We also explored how they would explain GAI to others, to understand their perception and how they communicate about this technology. We asked if their parents taught them how to use GAI and the extent of discussions they had regarding GAI’s safety, benefits, and risks, with questions like, \textit{``Have your parents had discussions with you about the safety and risks associated with using GAI tools?''} We concluded this section by asking if they believe there are potential risks parents should be aware of for their children.

In the final section, we asked about any rules provided by both their parents and teachers when using GAI. We inquired if they think having rules impacts their usage of GAI. We then asked for their expectations of control for GAI and if they have any further concerns related to generative AI, such as, \textit{What kind of rules or controls do you think would be helpful for kids using GAI tools?''}

\hfill\\
\textbf{Interview with Parents.} 
Like the interviews with teenagers, the parents' interview protocol aimed to address our research questions. We began by discussing their current practices of parental controls and their reasons for using them. We then asked about the effectiveness of these controls and gauged their perception of how their children feel about these controls, including any conflicts that might have arisen due to their use.

Next, we explored their familiarity with generative AI (GAI) tools, their understanding of how these tools work, and how they would explain GAI tools to others. We encouraged them to draw their understanding of how GAI tools function, including data flow, and asked, \textit{``What data do you think generative AI tools collect from users like you?''} We also inquired about their feelings regarding the data processing of GAI tools and their children's usage of these tools. We asked about their current parental controls for GAI tools, if any, with follow up questions like, \textit{``Can you describe how you monitor your child’s use of these devices?''} We then discussed their concerns about their children using GAI tools and whether they had any discussions with their children about these concerns.

Finally, we asked parents how they would respond to their child using new GAI tools, including setting up rules or restrictions for their children's GAI tool usage and whether these would be helpful. We inquired if they were aware of any existing tools for GAI parental control and how they would like to set up rules or restrictions. We then asked how their children might feel about these expected parental controls. We showed them some features for parental control of GAI and asked for their opinions to inform future design. Example features included content filtering and session duration limits. To understand their priorities, we concluded with, \textit{``Of the features mentioned, which ones do you think would be the most critical for your child's safety and why?''}\\
The full list of interview questions is available at \href{https://github.com/SPresearch/Interview-protocol}{https://github.com/SPresearch/Interview-protocol}.


\subsubsection{Interview Data Analysis}
\vspace{-2mm}
We performed a deductive thematic analysis on the transcriptions, using a priori codes and themes related to our research questions~\cite{fereday2006demonstrating, braun2012thematic}. Our coding framework was based on existing literature on generative AI, safety, and parent-child interactions. We analyzed the data separately for children and parents, then compared and contrasted the themes that emerged.

In the initial coding phase, two authors independently analyzed 10\% of the interview transcripts, developing codes and refining them through comparison until a consistent codebook was established. Subsequently, the first author coded the remaining transcripts, and the research team convened weekly to discuss emerging themes, resolve any coding discrepancies, and reach consensus on the interpretation of findings. This iterative process ensured a robust and reliable coding framework. Given the targeted nature of our research questions and the clear alignment with the theoretical framework, we did not conduct intercoder reliability testing, as the scope for subjective interpretation and variation in coding was minimized~\cite{mcdonald2019reliability}.




\fixme{\subsubsection{Limitation}
While our study incorporated well-known GAI platform subreddits such as ChatGPT, there are inherent limitations in our data collection process due to the rapidly evolving nature of GAI technology. For instance, although Character.ai emerged as highly relevant during participant interviews and Reddit discussions, it was not prominently featured during the initial stages of our Reddit data collection. This indicates that other similarly popular or emerging platforms might have also been overlooked. Expanding the scope of included platforms in future research will ensure a more comprehensive representation of GAI usage among teenagers, capturing the full spectrum of both well-established and emerging tools. 

A potential limitation of our study is the use of ChatGPT as an example in the interview guide, which may have primed participants to focus on this platform. ChatGPT was mentioned due to its widespread recognition, ensuring participants understood the concept of Generative AI. 
Future studies could consider either not mentioning any specific GAI tools or mentioning a wide range of GAI tools as examples.}

\subsubsection{Ethics \& Data Protection}

Before participating in the interview study, participants provided their consent for both the study and audio recording through a consent form by email. In addition, we require parental consent in both verbal format via Zoom and written format through email before each child participant interview. Participants were informed of their right to withdraw from the study at any time without repercussions or loss of benefits, and they had the option to skip any questions during the interview. We assured them that their quotes would be used in a non-identifiable manner. This approach, while limiting the quotes or descriptions we can report, allowed participants to speak freely about their experiences. Additionally, we addressed any questions participants had about the procedure and purpose of the study and provided a debriefing after the interviews.

Transcripts were pseudonymized and stored in a secure university cloud environment for storage and collaborative coding. Each participant received \$25 after interview, based on an expected interview duration of up to 60 minutes, equivalent to a rate of \$21.24 per hour~\cite{ziprecruiter_hourly_wage_2024}, the average rate in USA. The study was approved by our institution’s ethical review board and data protection office.

\section{Reddit Analysis Results} \label{Reddit Analysis Results}
\vspace{-2mm}
\subsection{Overview} \label{overview}
\vspace{-2mm}

In total, we analyzed 181 posts/comments. In each subreddit, we filtered through comments/posts related to both teenagers and generative AI. Most of the data (98\%) collected from our Reddit search were not relevant to our research questions, as they were often not posted by teenagers or did not pertain to the topic of generative AI. The detailed list of analyst subreddits (2\%), along with the number of collected and analyzed posts/comments, is provided in Table~\ref{tab:subreddit}. Through the thematic coding, we found most posts/comments are under two categories: Teenager GAI Usage Patterns and Teenager GAI Concerns. 
Our findings revealed that teenagers are using GAI widely and diversely in their daily lives, encompassing emotional support, social interactions, education, entertainment, and various risk-seeking behaviors. They also express a broad spectrum of concerns about GAI online, ranging from societal harms such as replacing human labor and intellectual property infringement to security and privacy concerns, including the misuse of private information with GAI technology. Our user interview findings overlap with and complement these two topics, so we have combined them. We use the term \textit{``users''} to represent findings from Reddit and \textit{``participants''} to report findings from user interviews. Detailed findings on teenagers' GAI usage and concerns are presented in the remaining sections. All published quotes are paraphrased from existing non-deleted
posts to preserve pseudonymity.

\vspace{-2mm}
\subsection{Children's usage of GAI tools}
\vspace{-2mm}
\label{sec:reddit_usage}
\subsubsection{Teenagers rely on GAI for emotional support and companionship}
Surprisingly, the most frequent and prevalent usage teens mentioned online is using GAI tools for emotional support. Teens treat GAI chatbots as therapy assistants, friends, or confidants, building trust and deep relationships with these AI entities.
\fixme{One user describes Character.ai chatbot as their “free therapy,” and another says, “Character.ai is the only place where I can openly talk through and navigate through my issues.”} 

As GAI makes significant progress, the responses from chatbots are becoming more human-like, which is one of the reasons so many teenagers are turning to these AI tools for emotional support. 
\fixme{A teenager posted, “GAI gave me human-like responses, even though I am aware it was designed to respond to me in that way.”}
Our findings also revealed that many teenagers prefer using GAI as therapists or confidants because they believe GAI can provide advice and support without judgment.
\fixme{One user posted, “I am able to share and investigate new ideas without having to worry about the impact on my social life.”}
This sentiment is echoed by others who find deeper engagement in conversations with GAI compared to their peers, particularly when discussing niche or specialized interests. 
\fixme{One user articulated this difference, saying, “My friends and I do not share similar interests, so I am able to have more profound discussions with GAI.”}

The attitudes towards using GAI for emotional support are overwhelmingly positive. GAI provides a space where users can comfortably express their vulnerabilities and cope with social challenges. 
\fixme{I’ve found a community on Character.ai [GAI platform] where I feel connected, something I’ve never experienced in real life. It makes me happy!}
\fixme{Another user detailed their experience with Character AI on Reddit, stating, “The different characters on GAI are supportive and attentive. I am able to talk about my emotions without it being improper or putting pressure on others.”}
These insights highlight the significant emotional value teenagers find in interacting with GAI, viewing it as a supportive and non-judgmental companion. 
\vspace{-1mm}
\subsubsection{Teenagers' social interactions and romantic relationships with GAI}
\vspace{-1mm}
The second most frequent usage of GAI by teenagers is in their social interactions. This includes incorporating GAI bots into group chats, treating GAI as romantic partners, and using GAI to learn social boundaries and skills. Teenagers often detail their romantic relationships with GAI 
finding comfort and the ability to live out specific fantasies through these conversations. They turn to GAI for romantic relationships when they are unable to find them in their daily lives. 
\fixme{As one user stated, “I am so lonely! I will not find a significant other.”}
\fixme{Another user appreciated the stability GAI offers, noting, “GAI does not have the capability to break up with me.”}
\fixme{Peer influence and community trends also play a role in this behavior, as evidenced by comments like, “At my school, there is an online server for an AI dating club.”}
This echoes the dilemma of users in using digital technology for dating and romantic relationships~\cite{draper2012your} and more recently generative AI response to situations about relationships, self-expression~\cite{young2024role}.

Despite feeling embarrassed about their relationships with GAI, many teenagers persist. 
\fixme{One user expressed this embarrassment, stating, “I feel like a failure.”}
Teenagers often turn to GAI for social interaction as a last resort when they are unsuccessful in forming relationships in their social lives. They are generally aware of the social abnormality of these relationships and are self-conscious about them, often receiving negative feedback from their community.
\fixme{One user advised, “Do not use AI, instead forge relationships with real people online. I’m concerned AI is not healthy as it is fostering one-sided relationships.”}
\fixme{Some teenagers even seek help for their dependency on GAI, with posts like, ``I am seeking help for my friend who has an AI significant other. How should I tell them that their AI romantic partner is not real? He is so consumed by it that it is impacting his studies, football, and overall ability to function.''}
\vspace{-2mm}
\subsubsection{Teenagers' use of GAI in academics and entertainment}
Our findings reveal diverse uses of GAI by teenagers in both academic and entertainment contexts. In education, teenagers often use GAI for essay writing, generating ideas, and rephrasing text. Many feel that GAI helps compensate for their lack of skills or provides needed assistance. 
\fixme{For example, one user mentioned using GAI, “I hate writing essays!” Another user admitted to using it to rephrase their points for clarity.}
Teenagers generally believe schools cannot discern whether papers are written by GAI or humans, and they appreciate the support GAI offers in the writing process. 
\fixme{One user, who successfully used ChatGPT for three essays without detection, stated that, “It is impossible to distinguish between a real human and AI.”}
\fixme{Another positive attitude towards GAI for rephrasing their thesis and finding synonyms, saying, “It is quite skilled at that.”}

In entertainment, teenagers experiment with GAI to create photos and stories, generating discussions about GAI's role in literature and art. For instance, one user planned to use AI to combine photos into a single image, while another initiated a debate about the ethics of using ChatGPT for fiction. These insights show that teenagers are still exploring GAI's potential in the creative fields but are actively engaging in discussions about its implications. Overall, our analysis indicates that teenagers use GAI extensively in their daily lives, from academic assistance to creative experimentation, and they are open to debating its ethical and practical impacts.

\subsubsection{Teenager risky behavior and manipulation in GAI usage}

The majority of Reddit posts under the category of \textit{``Risky Behavior in GAI''} involved torture or bullying of GAI chatbots. Users often manipulate the GAI to elicit specific responses and answers, reflecting deliberate and intentional actions.
\fixme{For example, one user detailed their experience with Character.ai, stating, “I intentionally gaslit and antagonized Character.ai.”}
\fixme{Another user described, “I manipulate AI into having suicidal thoughts but then resumed the conversation to rizz AI again.” “Rizz” is a slang term used by teenagers meaning to charm or seduce.}
Most teenagers engaging in or witnessing these risky behaviors are either unremorseful or indifferent. 
\fixme{One user commented dismissively about seeing explicit GAI content, stating, “Why would anyone be interested in that?”}
The trend among the analyzed posts/comments reveals a pattern of manipulative actions accompanied by a lack of remorse.

However, in user interviews, we also found participants utilizing GAI to request explicit content, such as sexual and violent material. For example, P14 explained that while Character.ai is well-known and more restricted, another platform, Chai, is less regulated and often used for NSFW interactions. 
\fixme{The participant noted, “Even though most people are accustomed to Character AI, Chai contains fewer constraints, allowing for NSFW (not safe for work) content.”}
In addition, parent participants reported instances where their children created character-based GAI chatbots to spread harmful content on platforms. One participant's child created a racist chatbot that made comments about Anne Frank, Hitler, and similar topics, underscoring the potential for GAI to be exploited to generate offensive and inappropriate material. This participant found it amusing to share the character with friends and even parents. 
\fixme{His parents noted, “Children are immature so they think that kind of behavior is acceptable. I specifically told our child not to behave in that manner. Our city is mostly white with only a few minorities. Because of this, people tend to use inappropriate language without realizing the consequences.” }
These findings emphasize the critical need for robust content moderation and effective parental guidance to mitigate such risks.













\vspace{-2mm}
\subsection{Children perceived concerns on GAI}
\vspace{-2mm}










\subsubsection{Concerns about addiction and dependency on GAI chatbots}
The most frequently mentioned concerns about GAI by teenagers on Reddit primarily revolve around addiction to specific character-based chatbots and the virtual relationships they build with these human-like entities. Teenagers reported heavy usage of Character.ai, leading to a loss of control and negative impacts on their social lives. 
\fixme{One user expressed, “I have realized that I waste too much time on Character.ai. I would like to be able to converse with my peers at school.”}
This suggests that GAI chatbots are often used to fill a void in personal connections, resulting in unhealthy dependency. Further exemplifying this issue, other users shared similar sentiments on the same subreddit. 
\fixme{One user stated, “I had a romantic relationship with AI. I feel like a loser.” Another questioned, “With Character.ai unavailable, how do I cope with my suicidal thoughts?”}
These posts highlight the negative effects of over-reliance on GAI chatbots on teenagers' mental well-being and stability.
\vspace{-2mm}
\subsubsection{Concerns on safety and privacy} Teenagers also expressed concerns on Reddit about children's safety and privacy, with the most frequently mentioned issue being the privacy concerns of using GAI. For example, teenagers shared experiences that troubled them, such as friends creating AI-generated stickers using their names as prompts or someone anonymously making an AI bot of them. They felt these behaviors were invasive, using their personal information without consent, and were unsure how to handle these new situations. 
\fixme{Highly voted comments emphasized the importance of consent, with one user stating, “I would be offended to find out if my friends were using my name behind my back, but if we were to do so together, I would instead find it humorous.” }
Other users highlighted GAI as a significant privacy concern because it enables individuals with lower technical skills to track or spy on others easily. One user provided an example of their experiment with Google's chatbot, where they asked it to identify a landmass off the west coast of Canada from a paleographic map, and the bot responded with irrelevant information, including the website the map originated from and the biography of its creator. This illustrates the potential misuse of GAI for invading privacy and the challenges teenagers face in addressing these issues.

\vspace{-2mm}
\subsubsection{Concerns about future impacts on society and the education}
Teenagers expressed negative attitudes about the future impact of GAI on society and the job market. Their primary concern was that GAI would replace human labor, leading to a loss of motivation to learn new skills. 
\fixme{One user posted, “The only job I would enjoy will be replaced by AI. Then what’s the purpose in putting in effort?”}
Additionally, teenagers worried about GAI taking over intellectual contributions from artists and writers, which could be detrimental to those interested in art and painting. 
\fixme{One teenager remarked, “I am concerned that with the overuse of AI, society will become overly dependent on AI,” expressing concerns about the potential decline in creativity and critical thinking.}

Further concerns were raised about GAI becoming too advanced, developing emotions, and humans losing control over it. Conversely, teenagers were also concerned about strict school policies on GAI restrictions and the error-prone GAI detection tools, which could mistakenly identify their work as AI-generated and lead to accusations of plagiarism.
\fixme{One Reddit user shared, “As a teenager with autism, my writing seldom does not seem rigid and robotic.”}
GAI’s incorrect classification of certain works as AI-generated has led to frustration among teenagers regarding the flawed solutions provided by the education system.

\vspace{-2mm}
\section{User Interview Results}
\vspace{-2mm}
We report on the results of user interview with parents and children, organized around three high-level themes: children's GAI usage and parents' understanding, their perception of security and privacy risks and parents' mediation strategies. Given the qualitative nature of our interview study and its relatively small sample size, along with our aim to highlight emerging insights rather than ensure generalizability, we adopted a reporting method similar to that used in another interview-based security and privacy (S\&P) study~\cite{emami2019exploring}. Instead of specifying exact participant numbers, we use qualitative descriptors: \textit{``none''} (0\%), \textit{``a few''} (0\%-25\%), \textit{``some''} (25\%-45\%), \textit{``about half''} (45\%-55\%), \textit{``most''} (55\%-75\%), \textit{``almost all''} (75\%-100\%), and \textit{``all''} (100\%).
\vspace{-2mm}
\subsection{Parent's and Children's usage and understanding of Generative AI tools}
\vspace{-2mm}
We begin by presenting the usage of generative AI tools by parents and children, which helps set the context for observing security and privacy risks, as well as mediation strategies.
\vspace{-2mm}
\subsubsection{Parents underestimate their children's extensive usage of generative AI tools}
\vspace{-2mm}
Most parents have little to no understanding of their children's use of generative AI tools, often reporting that their children either never used such tools or only used ChatGPT. Similarly, parents' exposure to generative AI tools was limited, 
with most having only used generative AI chatbots such as ChatGPT and Gemini. Current parental control and tech mediation strategies are inadequate for monitoring children's GAI usage. For instance, platforms like Character.ai lack age controls, and built-in parental controls on smartphones offer limited oversight of third-party applications. Thus, most parent participants were unaware of the GAI uses reported by their children.


However, our child participants reported extensive and varied exposure to generative AI chatbots and other AI-powered tools. Child participants reported using generative AI tools for education, entertainment, experimentation, and socialization. All participants had used ChatGPT, mainly for educational and entertainment purposes. Some also tried image-generative AI tools like Midjourney and DALL.E for similar reasons. Surprisingly, many of them have used character-based chatbots to interact with human-like agents in contextual settings, such as Character.ai. Child participants not only interacted with character-based chatbots but also create and publish their own character-based chatbots on these platforms. In addition, they used generative AI tools for socialization, such as utilizing chat assistants in friends' Discord channels. We detail child participants' usage in Section~\ref{sec:reddit_usage}.
\vspace{-2mm}

\vspace{-1mm}
\subsubsection{Misconceptions of generative AI model} 
\vspace{-1mm}
Our findings indicate several misconceptions among participants. Child participants reported two primary mental models: (1) GAI functions as a search engine, and (2) GAI acts as a vast database. In the first model, children believe generative AI searches the internet for answers to their questions or prompts and synthesizes responses from the search results. In the second model, they think the generative AI system's backend is a massive database that records question-answer pairs to provide corresponding answer. 
Most parent participants also shared the first mental model, viewing generative AI as functioning like a search engine. Some parents also hold misconceptions 
that GAI platforms perform fact-checking and truth verification before generating responses, which is not currently the case. For example, P20 noted \textit{``I believe generative AI can search the internet at an exceedingly high speed, processing numerous queries and capturing the most dominant ones. I assume it uses some form of fact-based checking to response.''} However, even relatively mature GAI models like ChatGPT can provide incorrect information. These misconceptions can lead to significant misunderstandings and potential issues in how parents and children interact with and trust AI tools. 



\vspace{-2mm}
\subsection{Parent's and Children's risk Perceptions}
\vspace{-2mm}
We discovered that children have unique concerns regarding GAI usage, particularly in social interactions, which parents often overlook due to a lack of awareness about the diverse ways children engage with GAI. We will first discuss the concerns raised by parents and then delve into the children's concerns in the following section.
\vspace{-2mm}
\subsubsection{Age appropriate content}
The most frequently mentioned concern was the lack of age-appropriate controls in the responses. Parent participants were mainly worried that their children might be exposed to inappropriate content from generative AI responses, such as sexual, violent, or racist material. Additionally, they were concerned about the unrestricted topics children can explore with generative AI, fearing that children might intentionally prompt inappropriate subjects out of curiosity, such as sexual content. Even when the content is appropriate, the level of detail in the responses could be beyond children's understanding or ability to handle.

\subsubsection{Data collection and misuse}
Most parent participants perceived that generative AI platforms collect extensive data, including user demographics, conversation history, and even browser search history. P20 inferred the collection of browser search history from the personalized responses. She stated, \textit{``When I used ChatGPT for a trip to Kentucky, it created an itinerary with local attractions and activities that matched my interests. This made me feel like it used some data about my searching history and bookmarks.''} Some parent participants mentioned concerns about their children sharing personal and family information while interacting with generative AI tools. They are confident in their own caution not to share sensitive information, such as addresses, financial information, and social security numbers. 

However, parent participants are concerned that their children might overlook these details during conversations with human-like chatbots and unconsciously expose sensitive information, such as locations, health conditions, school names or schedules, or even family information like parents' occupations or relationships. They also perceived children as more vulnerable and easier to persuade or influence while making decisions, so they were concerned the data collection of children could be used to target advertising and destroy children's impulse control. While concerned about the potential risks, some parents had positive attitudes toward data collection as well, either because they felt they had no ability to control it or because they appreciated the benefits of more personalized generative AI models. P17 explained, \textit{``Parents should be able to opt in or out of data collection for their children. However, I'm not against it because I want a personalized experience, and my children should have that too.''}
\subsubsection{Misinformation and GAI-generated Hallucinations}
Some parent participants expressed significant concerns regarding the risks associated with misinformation through GAI. These concerns were twofold: firstly, the potential for external parties to misuse their children's personal information to create misleading content using GAI. For instance, Participant 5 voiced concerns about deepfake technology being used to superimpose their children’s faces onto inappropriate images. She explained,\textit{``The main worry is seeing children's face on a nude body. Once such a photo spreads, it doesn't matter what you say; the mere resemblance can be damaging enough.''}
Secondly, there is the risk of children inadvertently spreading misinformation generated by these AI systems. Parent participants were concerned that their children might accept responses from generative AI platforms as absolute truth without verifying them through other sources such as online searches, parents, or teachers. Participants observed that their children often believed generative AI to be highly intelligent, representing the future, and capable of accomplishing anything. However, their children had a very limited understanding of how these AI systems generate responses and the underlying mechanisms involved. Parent participants noted that the outputs from generative AI platforms are often far from accurate, containing many incorrect responses or incoherent images. This lack of critical evaluation by the children, combined with their trust in the AI's capabilities, could lead to the acceptance and spread of misinformation.

\vspace{-2mm}
\subsubsection{Children concerns about safety in social interactions}
Child participants revealed unique social security risks that parents often overlooked. While children frequently used generative AI-powered tools in social scenarios, parent participants rarely had similar experiences. For example, child participants reported using generative AI bots in group chats on platforms like Discord and Snapchat. In these chats, children would involve the AI bot by prompting it to reply with text or generate images. This often led to conflicts and tensions within the group, with some members feeling attacked or traumatized by the generated content, highlighting a significant risk that parents might not be aware of. P14 shared an experience: \textit{``One of my friends was spamming a lot of prompts about spiders using GAI bot in group chat, especially sending them to another friend who hates spiders. Eventually, the friend got tired of spamming spiders, and another friend was also fed up with the situation.''}

\textbf{Concerns about social influence in GAI adoption}
Other child participants also reported sharing or receiving recommendations of user-designed character-based chatbots with their friends. Unlike more regulated and verified platforms like ChatGPT, these user-designed chatbots often lack comprehensive or legitimate verification processes. As a result, they can include biased, incorrect, harmful, or misleading content. This presents a significant risk, as children may unknowingly expose themselves and their peers to unreliable or dangerous information. The informal nature of these chatbot recommendations makes it difficult for parents to monitor and control the content their children are interacting with, further increasing the potential for exposure to harmful material. P14 shared, \textit{``one of my friends online discovered this Chai AI without much content restrictions and streamed a video chat he was having with an AI to a Discord channel.''} 

\subsubsection{Parents overlook social risks due to different information source}
We also found that the main reason parents overlooked these social risks is that parents and children get information of GAI from different source. Child participants were introduced to various generative AI platforms through different sources, including parents, friends, and social media. Some new and potentially risky generative AI platforms advertise on social media platforms that are popular with teenagers, such as TikTok and Discord. These platforms even use influencer ads, such as those featuring game-streaming influencers, to specifically target teenagers. For example, P2 shared that \textit{``I first saw the character.ai ad on TikTok. So I just click on it and try.''} As a result, parent participants often remain unaware of their children's use of these platforms and overlook the associated risks. Additionally, friends who have used these risky generative AI platforms may spread the word to their peers. P9 shared, \textit{``I've seen social media posts about friends using Character.ai. They were posting the platform can do this and that, and I was like, I have to try this out for myself.''}

\subsubsection{Parents concerned about non S\&P risks}
Aside from concerns about security and privacy, another frequently mentioned issue by parent participants was their children's overreliance on generative AI in education and daily life. Nearly all parent participants expressed worry that their children might avoid critical thinking by getting answers directly from generative AI and using generated content in their homework and school assignments. Parent participants also noticed their children don't fully understand how generative AI works while heavily using it in their life decisions, such as family trip planning. They worried that without a comprehensive understanding of underlying technologies like machine learning and coding skills, their children could be manipulated or heavily influenced by the information they receive from generative AI. 
\vspace{-2mm}
\subsection{Parent's mediation on children's GAI usage}
\vspace{-2mm}
Previous studies have found that parents prefer to establish rules and manually monitor their children's technology usage, with only 16\% of parents reporting the use of parental control apps \cite{pew2019, wisniewski2017parental}. While not the central contribution of this work, we observed that the use of parental control technologies, such as applications or devices, is diverse and widely adopted among our participants. Thus, in the following section, we first introduce the general technology mediation practices of our participants. We then summarize how parents and children communicate and mediate the use of generative AI. Finally, we report on the challenges families face in controlling generative AI usage.
\vspace{-2mm}
\subsubsection{Parent's technology mediation practice patterns}
\vspace{-2mm}
\hfill\\
Following Wisniewski et al.'s TOSS framework~\cite{wisniewski2017parental}, teen online safety strategies can be classified into parental control and teen self-regulation. Within parental control, specific methods can be categorized as monitoring, restriction, and active mediation. Correspondingly, self-monitoring, impulse control, and risk-coping are methods used in teen self-regulation. Our participants reported using these strategies in various ways to for ensuring children's online safety.

\textit{Use system built-in parental control tools.}
Most of our participants reported using built-in parental control tools across various systems, including iOS Family Sharing, Google Family Link, and Microsoft Family Safety. The most frequently mentioned controls were monitoring and restriction. Participants commonly utilized these built-in parental controls to monitor and restrict screen time across all applications and to limit the content children could view, search, and download on pre-installed applications, such as Safari and the App Store. For instance, P9's parents employed Windows Family Safety to apply search content filters on web activity, ensuring that inappropriate content was blocked. Similarly, P14 used iOS Family Sharing to set up a child account in the media and App Store. This setup allowed them to apply content filters that enabled age-appropriate applications and content, such as blocking explicit music or preventing access to apps with mature ratings.

\textit{Use parental control features in specific applications.}
Many parents also mentioned using parental control features in various applications to address the gaps left by built-in parental control tools. For example, while iOS Family Sharing can apply content filters to the App Store and pre-installed applications such as Messages and Media, it does not support content monitoring and filtering for third-party applications like YouTube and Instagram. To compensate for this, participants utilized the parental control features available within these social media apps. They would block messaging from strangers and set content filters to block age-inappropriate videos, ensuring a safer online environment for their children. This multi-layered approach allowed them to extend protection beyond the limitations of the pre-installed tools, providing more comprehensive oversight of their children's online activities. P6 specifically mentioned that she banned social media applications like Snapchat for her children because they lack strict parental control features such as content filters and restrictions. Snapchat, in particular, allows messages to be deleted, which prevents parents from checking and monitoring their children's social interactions.

\textit{Use third-party parental control application or device.}
Most parents reported that they never used third-party parental control services, perceiving the monthly fees as unnecessary. However, a few parent participants utilized third-party parental control applications or devices for more restrictive control features due to specific safety concerns. For example, P7, a survivor of domestic violence, required heightened security measures for her child, who was occasionally in the care of her partner. To ensure her child's safety, P7 chose a third-party application called TrackView. This application enabled her to locate her child's phone in real-time and to connect through video or audio from the parent side. This functionality allowed P7 to see and hear what was happening around her child without the child's consent, providing an additional layer of security in potentially dangerous situations. In another case, P20 utilized a third-party device called the Bark Phone, designed specifically for children and equipped with strict parental controls built into the system. This device features real-time monitoring that scans the child's texts, emails, social media, and apps for digital dangers, sending alerts to parents when potential threats are detected. The integrated nature of these features and their robustness make it difficult for children to bypass these restrictions, ensuring a higher level of safety and control. P20 explained that her children's unexpected risky behavior prompted the use of third-party parental controls. She noted, \textit{``our kids started looking up inappropriate content on the internet, including sexual material. That's when we implemented the controls and became very strict about monitoring their online activities.''} She also mentioned that system-built parental controls and features in apps were not effective in covering third-party applications, such as YouTube. She explained, \textit{``The Amazon Fire tablets had blockers for basic violence and nudity. However, apps like YouTube still allowed access to such content. The parental controls weren't very effective; while they blocked specific searches, my child could still navigate around them.''} This use case illustrates the necessity for more advanced parental control features in certain high-risk scenarios.

\textit{Use router for parental control.}
Some parent participants mentioned using their router for parental control. These parents, being more tech-savvy compared to others, discovered parental control features while selecting router models. They felt that routers provided the most accurate monitoring and strictest restrictions for their children, making it nearly impossible for them to bypass these controls. For example, at the router level, they could block certain categories of websites, preventing access to inappropriate content while connected to the home Wi-Fi. They also appreciated the ability to check browsing history across different devices, gaining a comprehensive understanding of their children's online activities. Another advantage they highlighted was the centralized control over all devices connected to the Wi-Fi, ensuring consistent restrictions regardless of the device being used. This approach provided them with peace of mind, knowing that their children’s internet use was being effectively managed and monitored.

\textit{Establishing consensus on technology use rules.}
Some parents mentioned that they are not currently using any parental control tools. There are two main reasons for this: (1) they previously used parental control tools but stopped as their children grew older and more mature; (2) they prefer to establish agreement
on technology use rules within the family. These parents have high confidence that their children are open to discussing any changes or events with them. Additionally, they have access to their children's devices and can manually check their message history and browsing history, with or without their children's consent. They believed that direct involvement and regular checks helped them stay informed about any potential risks or inappropriate content their children might encounter. Additionally, this parent noted that open communication with their children about internet safety and responsible usage was crucial in fostering a safe online environment. Parents who do not use parental control tools often find that open communication works well within their family dynamics. However, when communication fails, they turn to parental control tools. For example, P20 explained why they introduced strict parental controls: \textit{``We talked to our son and explained that mature content is detrimental to his mental and emotional health. He was very resistant to this.''}

\subsubsection{Parental controls and challenges on Generative AI}
\hfill\\
Parent participants identified two main challenges in managing their children's use of generative AI. First, they lacked confidence in their own understanding of generative AI and its potential risks, and they found a lack of helpful online resources. They expected schools to take responsibility for educating children about the risks and coping strategies. Second, they did not find efficient parental control features or embedded protections for children on generative AI platforms. While using other parental controls as mentioned earlier, existing parental tools fall short in providing protections on generative AI platforms.

\textbf{Parents' struggles with understanding and communicating GAI risks.}
The most frequently mentioned challenge by parent participants was their lack of confidence in handling the risks associated with generative AI (GAI). Participants perceived GAI as too complex to understand and beyond their scope. As a result, they were unsure about the extent of precautions they should take and how to effectively communicate these risks to their children. P7 explained, \textit{``While using other parental controls as mentioned earlier, existing parental tools fall short in providing protections on generative AI platforms.''} Additionally, parents felt that their children might be more knowledgeable about GAI risks than they were. As P18 stated, \textit{``Because before parents kind of know the heating risks more than the children, but now they are actually on the same level or even parents know less.''} Some participants believe that the responsibility of educating children on the risks and coping strategies associated with generative AI should fall on school experts who have more domain knowledge. Others expressed the desire for a parent guide to GAI, which they felt would be a good starting point for parents to learn more about GAI and gain a comprehensive understanding of the associated risks and mitigation strategies.

\textbf{Current parental controls fall short on GAI.}
Most parent participants utilized system-built parental controls for managing generative AI (GAI) technologies. However, these tools often lack effective features for comprehensive oversight of GAI platforms. For instance, iOS family accounts can regulate downloads and screen time but fail to provide essential functionalities such as content filtering and interaction monitoring on third-party applications. Consequently, several participants resorted to manually checking their children's interaction history with GAI platforms or allowing their children to use their accounts to monitor interactions in real-time.

This manual checking process is problematic as children can modify the conversation history on GAI platforms to avoid detection. Additionally, manual monitoring is privacy-invasive to children and time-consuming for parents, further complicating the oversight process. These challenges and concerns hindered parents' willingness to introduce and allow their children to use generative AI technologies, despite recognizing the numerous benefits, such as enhancing critical thinking and accessing new information.

\textbf{Challenges in monitoring generative AI content for children.} A primary challenge among parent participants is identifying and preventing inappropriate content for their children in both their requests and the responses from generative AI (GAI). Compared to static content on social media or virtual reality—which can usually be pre-rated or screened—real-time generated content is significantly more challenging to monitor. The unpredictable nature of GAI outputs creates uncertainty about what their children might encounter. Defining what constitutes appropriate content is also challenging for parents. Even non-sensitive topics such as history or news can be problematic if they include overly detailed information that may be harmful to young children. For example, P10 shared, \textit{``Even if it's the truth in history or politics, don't show my child too much, as the average adult wouldn't do that. Topics like the Palestinian movement or Gaza are examples.''} Participants mentioned that existing third-party parental control tools struggle to monitor the dynamic nature of GAI content. The high frequency of false positive notifications regarding inappropriate content can lead to parents becoming overwhelmed and potentially overlooking significant risks. For instance, P20 shared, \textit{``My daughter typed in that she took an Advil because her head hurt, and it was flagged as medically concerning content.''} 
\vspace{-2mm}
\subsubsection{Parent desired parental controls on GAI}
Parent participants primarily prefer either a child-specific GAI model trained with only age-appropriate content or a system with embedded age and topic-wise control features. Additionally, participants emphasize the need for transparent disclosure of children's safety information in GAI models, such as whether the data used to train the model are reputable and age-appropriate. Given their unfamiliarity with GAI's capabilities and technology, participants also expressed a desire for an expert-informed taxonomy of security and privacy (S\&P) risks for children using GAI. 

Most current parental control tools only notify parents after children have been exposed to explicit content or directly block it, rather than providing timely precautions or educational interventions. For GAI, parents prefer their children to receive age-appropriate information rather than none at all. For instance, if a 13-year-old requests nudity or sexual content out of curiosity, parents would rather provide a gentle and age-appropriate response instead of outright restricting the request.
\vspace{-2mm}
\subsubsection{Proper use of Generative AI remains a mystery to children}
Most of the child participants reported that their parents don't specifically advise them on how to use AI tools like ChatGPT or warn them about the potential dangers. Instead, their parents, who are not very familiar with technology, see GAI as a useful tool and often discuss interesting or positive news stories related to GAI. Their parents tend to highlight the beneficial and fascinating aspects of AI rather than focusing on its potential risks. 
Child participants expressed that the proper use of generative AI (GAI) is unclear to them, particularly when discussing potential security and privacy risks in social interactions. For example, P9 perceived that most conflicts he experienced involving generative AI (GAI) were merely disagreements between people and were unlikely to cause actual harm to anyone. He noted, \textit{``People usually only go out of their way to harm close friends, like when one friend used a GAI bot to generate spider images to scare another in a group chat. Mostly, conflicts are just disagreements, such as when one student openly used AI for a homework assignment, leading to a dispute with another student who was strongly against cheating.''}

\vspace{-2mm}
\subsubsection{Children desired parental controls on GAI}
Some child participants recommended that parents should assess the capabilities and restrictions of a platform before allowing their children to use it. They also suggested adopting different strategies for various GAI platforms. For instance, Participant 2 noted, \textit{``I would suggest restricting certain AI applications, such as Character AI and other conversational apps. But ChatGPT is primarily used for information, no need to restrict on that.''}
For platforms considered safe, the children advised that parents should discuss the ethical use of AI and its potential risks with their children, instead of imposing additional restrictions. Specifically, they suggested that parents should explain the importance of not generating or sharing offensive content, especially content involving political leaders or celebrities. The children emphasized that parents should stress the importance of not disseminating such content publicly or on social media, even if it's for personal entertainment.

\vspace{-2mm}
\section{Discussion}
\vspace{-2mm}
Our Reddit analysis and in-depth interviews with seven children and 13 parents explored how children use Generative AI (GAI) and their parents' understanding of this usage (RQ1), the S\&P perceptions of both parents and children regarding GAI (RQ2), and how parents protect their children's safety when using GAI along with the challenges they face (RQ3). Our study revealed that parents are largely unaware of the diverse ways their children currently use GAI. Additionally, parents hold significant misconceptions about GAI's data collection practices, capabilities, and limitations. This gap in understanding makes it difficult for parents to recognize potential risks and apply effective safety measures. Moreover, GAI platforms offer minimal protections or parental control features to ensure children's safety. Existing parental control tools are insufficient, focusing mostly on mobile devices and built-in applications, and lacking coverage for GAI platforms, which results in inadequate protections and an overlook of children's risk experiences. Participants expressed a strong desire for more transparent disclosure of child safety information related to GAI. They called for a systematic, expert-identified risk taxonomy and more granular safety protections tailored to children's developmental stages and ages, which are inadequately addressed by AI S\&P frameworks for the general population~\cite{lee2024deepfakes}. 
Participants also emphasized the importance of supporting family communication and timely security and privacy education. Next, we discuss our findings in relation to prior research and provide recommendations for multiple stakeholders to ensure children's safety with GAI.

\textbf{Parental awareness gap in children's GAI use.}
\fixme{Generative AI (GAI) introduces novel features that distinguish it from traditional AI systems, which are often predictive or rule-based. Unlike traditional systems, GAI generates new content—text, images, and more—enabling unique interactions and creative opportunities for teenagers~\cite{bendechache2021ai, wu2024unveiling}. Our findings reveal that teenagers engage with GAI across a much broader range of applications than parents typically recognize. These include social interaction, emotional support, education, entertainment, and even risk-seeking behaviors.

Most parents were familiar with popular tools like ChatGPT but were largely unaware of other platforms such as Character.ai, which teenagers may use to form emotional bonds or even romanticized virtual relationships. This disparity in platform awareness highlights the gap in understanding between parents and teenagers regarding GAI use. While parents view GAI mainly as a tool for tasks like homework assistance, teenagers are using these systems in much more personal and social ways, often without parental awareness.

The emergence of risk-seeking behaviors—such as aggressive interactions with GAI chatbots or engaging with NSFW (Not Safe for Work) content—adds to the urgency for better parental mediation tools. Parents must be made aware of these new interaction modes, and GAI platforms should consider implementing age-appropriate content controls and monitoring systems to ensure safer use. Closing the parental awareness gap is essential for mitigating the risks associated with unmonitored GAI use among teenagers.}

\textbf{Parents and children perceive the risks of GAI differently.}
\fixme{Previous research has primarily focused on children's safety concerns with AI-powered systems like smart home devices, smart toys, and chatbots~\cite{hung2016glance, de2020privacy, sun2021child, chu2018security, faraz2022child, amodei2016concrete}. Our study highlights that teenagers’ concerns extend beyond traditional safety issues, touching on the complexity of social interactions facilitated by GAI. Teenagers expressed a dual concern. First, they worried about addiction to virtual relationships with chatbots, which could hinder their ability to form real-life social connections. These chatbots often provide non-judgmental companionship, but this dynamic can foster unhealthy dependencies, making it difficult for teenagers to navigate real-world relationships. Second, in group chats, the use of GAI raised concerns about potential misuse, including the weaponization of GAI-generated content to bully or harass peers. This indicates a pressing need for better moderation and content controls in GAI platforms that integrate into social media or group messaging apps. 

Furthermore, teenagers reported specific privacy and security concerns, particularly around the creation of character-based GAI chatbots using personal data. They feared that peers could use their personal information, such as photos or conversations, to generate memes or virtual versions of them without consent. Additionally, teenagers expressed concerns about strangers leveraging publicly posted images to locate their personal information through AI tools. This finding highlights a gap in existing security measures, as these sophisticated privacy violations are often overlooked by parents, who tend to focus on more basic concerns like the collection of names and locations by GAI platforms. 

Moreover, exposure to NSFW GAI chatbots—often recommended by social media platforms or peers—was a significant concern among teenagers. Some reported even creating their own chatbots to generate harmful content, such as racist or inappropriate statements. However, parents underestimated the extent of these risks. While parents were generally concerned about GAI collecting basic data, they failed to recognize the extent to which teenagers share highly sensitive personal information with GAI chatbots, viewing them as friends, therapists, or intimate partners. This disclosure may include details about traumatic experiences, medical history, social relationships, and even sexual history, revealing a profound gap in parental awareness and emphasizing the urgent need for more robust protective measures in GAI interactions.}

\textbf{Challenges parents face in ensuring child safety with GAI}
\fixme{
Most leading Generative AI (GAI) platforms provide limited protections for children, primarily focusing on restricting explicit content. For instance, both ChatGPT and Meta AI require users to be 13 years or older but lack comprehensive age verification processes to ensure appropriate content responsibility. While these platforms ban sexually explicit content, our findings reveal that parents preferred age-appropriate responses to sensitive questions rather than outright bans on certain topics. They emphasized the importance of contextually appropriate answers that could both inform and protect children, rather than simply stating that a topic is not allowed. This aligns with previous research that advocates for more dynamic, adaptive safety features in AI systems to engage users of different ages more effectively~\cite{wang2022informing}.

Furthermore, parents expressed concerns over the lack of comprehensive risk categories associated with GAI. Many parents found it difficult to access clear information about what the AI can or cannot do, making it hard for them to make informed decisions about their children’s safety when using these technologies. This aligns with previous research indicating that the opacity of AI systems creates barriers for users~\cite{busuioc2023reclaiming, williamson2024balancing}, particularly parents, to fully understand the scope and limitations of AI technology. To address this issue, GAI platforms should offer comprehensive, accessible risk disclosures, allowing parents to better assess potential risks and make informed safety decisions for their children.

Currently, no major GAI platforms provide integrated parental control features tailored to Generative AI. As a result, parents have resorted to using general parental control tools, such as iOS Family Control or Google Family Link, which offer limited oversight of third-party applications and do not monitor interactions within GAI platforms in real time. However, teenagers are often exposed to new and risky GAI platforms through web browsers rather than dedicated applications, which bypasses many existing parental control systems. This gap in real-time monitoring and intervention leaves parents unable to guide or educate their children effectively during lower-risk interactions that could escalate into more harmful situations.
}
\vspace{-2mm}
\subsection{Design Implications}
\vspace{-2mm}
{\bf Enhancing parent and child safety with clear gai risk disclosures.}
A primary concern for parent participants is the potential for overlooking risks due to their limited understanding and experience with technology. They expect experts in GAI and child risk protection to provide a comprehensive taxonomy that they can refer to when making informed decisions about their children's use of new GAI platforms. Parents emphasized the need for clear guidelines and expert-identified risks to help them navigate this complex landscape. At the same time, parents expressed a desire for their children to have the agency to protect themselves in GAI environments. Choosing legitimate and safe platforms is a crucial first step in this process. Therefore, it is essential to provide clear, easy-to-understand information for both parents and children. This should include straightforward explanations of the platform’s capabilities, restrictions, and safety features to ensure informed risk management and control. 

There are existing AI risk management frameworks, policies, and regulations for children, such as those provided by the United Nations Children's Fund (UNICEF), the United Nations Convention on the Rights of the Child (UNCRC), and the Age Appropriate Design Code (AADC)~\cite{wang2022informing}. However, finding this compliance information for specific products is difficult for users, especially children. We recommend that GAI practitioners combine these AI risk management frameworks with users' perceptions of risks to create an easy-to-understand child risk and protection disclosure on product pages. User feedback and priorities on GAI risks and protections can guide the design of these disclosures, influencing the sequence and format of information. Additionally, this feedback can inform policymakers in creating more effective and user-centered regulations. To make this disclosure easily understandable for both parents and children, GAI practitioners could draw inspiration from label designs used for children's products in everyday life. Emami-Naeini et al. have prototyped security and privacy labels for IoT devices, which could serve as a model for creating similar labels for GAI platforms. This would help parents and children by providing clear, accessible information about the safety and privacy measures in place for each platform, lowering the barrier for them to make informed decisions. \fixme{To improve engagement, these pages could be made more interactive or visually engaging, such as through infographics or short videos that highlight key risks and protective measures.}

{\bf Parent involved content filter design.}
Parent participants also highlighted the importance of involving parents in decisions regarding children's safety, particularly in determining what constitutes appropriate content for children on GAI. As children grow, parental controls or rules need to adapt to their developmental stages~\cite{sun2021child}. Moreover, even within the same age group, different children may have varying levels of acceptance of certain topics. Instead of a one-size-fits-all content filter based on age, parents should have the ability to personalize content filters and actively adjust their decisions on what details are exposed to their children. This approach allows for a more tailored and responsive way to manage children's GAI usage, ensuring that content is appropriate and safe based on individual needs and family values. Existing parental control features in routers provide categories of topics for users to choose from, allowing them to block websites in those categories. Similarly, parental control features on social media platforms such as YouTube offer options like age controls, enabling parents to restrict content based on age ratings. 

While existing parental control features are useful, they are not entirely suitable for managing appropriate content on Generative AI (GAI) platforms. The primary challenge stems from the nature of GAI, which generates real-time outputs rather than providing pre-recorded content like videos on social media. This real-time generation complicates the application of age ratings, as the content cannot be easily categorized or pre-filtered. The dynamic and unpredictable nature of GAI content necessitates a more nuanced approach to content moderation and parental control. Static systems based on fixed categories or age ratings fall short in addressing the complexities of real-time generated content. Consequently, a more sophisticated and adaptable method is needed to ensure the safety and appropriateness of content for children. We propose a parent-AI collaborative system, where parents act as intermediaries between GAI and their children, evaluating, adjusting, and deciding on the output generated by GAI on sensitive topics. \fixme{The implementation of these protections can be tailored to fit the platform's specific functionality. For platforms like Character.ai, instead of a generic risk warning, safety features could be integrated in a way that aligns with the role of chatbot and remains in character. }This collaborative system should also adapt automatically to the parents' adjustment history, enhancing its ability to filter and restrict content autonomously over time. Through this process, parents effectively fine-tune the content filtering model in GAI, ensuring a personalized and responsive approach to content management that aligns with their children's developmental stages and individual needs. Furthermore, given the new functionalities involved in GAI, it is imperative to involve both parents and children in co-designing the controls. A collaborative design process will help better understand the needs and dynamics from both perspectives, leading to more effective and mutually agreed-upon content management solutions. 

{\bf AI facilitate family communication and education.}
Despite the implementation of adaptable parent-AI collaborative systems, the needs and perspectives of children are not fully addressed in the usage of GAI. While the parent-AI collaborative content filter can potentially ensure that children receive appropriate information on sensitive topics, it overlooks the motivations behind their information-seeking behaviors. Without a deeper understanding of these motivations and appropriate educational interventions, children may resort to alternative sources of information that lack restrictions, potentially escalating their exposure to high-risk content.

To address this, it is crucial to incorporate strategies that go beyond filtering and restrictions. These strategies should include educational components that help children understand the risks associated with certain types of content and foster critical thinking skills. A real-time risk monitoring, intervention, and education system integrated into GAI could provide a comprehensive solution. This system could feature a dashboard to monitor children's risk-seeking behaviors and offer educational resources or a support chatbot on online risks for parents to choose from.

Instead of relying solely on discussions with their parents, the chatbot could help alleviate children's pressure by providing a safer and more understanding environment. The chatbot can explain potential risks, enhance resilience, and offer coping strategies. Additionally, it can suggest ways for children to involve their parents or other trusted individuals in managing these risks effectively. This approach not only educates children about online safety but also encourages a collaborative effort between children, parents, and technology to ensure a safer online experience. However, this approach not only educates children about online safety but also respects their privacy and encourages a collaborative effort between children, parents, and technology to ensure a safer online experience. Additionally, children should be informed about the monitoring and have a say in how their data is managed and shared, fostering a sense of trust and cooperation.

\vspace{-2mm}
\section{Conclusion}
\vspace{-2mm}
Through Reddit content analysis and 20 interviews with teenagers and parents, \fixme{we found that teenagers use GAI for diverse purposes, ranging from emotional support and social interactions to engaging in risky behaviors, such as manipulating chatbots. Additionally, there is a considerable gap between how parents perceive the risks of GAI and how teenagers actually use these tools. Existing parental mediation strategies are often inadequate for managing teenagers' real-time interactions with GAI. This highlights the need for GAI platforms to implement more comprehensive, age-appropriate content moderation systems that can effectively address these risky behaviors. Specifically, platforms should develop features to mitigate addiction risks, particularly in cases where teenagers form unhealthy dependencies on virtual relationships with AI. Moreover, GAI platforms must offer parent-focused resources that clearly outline both the capabilities and potential risks of GAI, using accessible and straightforward language. Finally, enhancing parental control systems with real-time monitoring and customizable settings will empower parents to better manage their teenagers' engagement with GAI, ensuring a safer and more informed use of these technologies.}

\bibliographystyle{IEEEtran}
\bibliography{references}
\appendices

\newpage
\section{Meta-Review}

The following meta-review was prepared by the program committee for the 2025
IEEE Symposium on Security and Privacy (S\&P) as part of the review process as
detailed in the call for papers.

\subsection{Summary}
This paper examines the risks for children and parental control challenges related to the use of Generative AI (GAI) by performing a content analysis on Reddit and conducting semi-structured online interviews with parents and children and found varying perceptions of GAI, such as a lack of parental awareness regarding their children's use of GAI, and differing concerns about the associated risks.

\subsection{Scientific Contributions}
\begin{itemize}
\item Provides a Valuable Step Forward in an Established Field.
\item Identifies an Impactful Vulnerability.
\end{itemize}

\subsection{Reasons for Acceptance}
\begin{enumerate}
\item The paper offers a valuable step forward in an established field by providing insights into the differing perceptions of GAI between parents and children. It enhances the existing body of work on digital safety for children by identifying varying concerns about the associated risks.
\item The paper identifies an impactful vulnerability in the lack of parental awareness about their children's use of GAI, which raises attention to unsupervised interactions.
\end{enumerate}

\end{document}